\documentclass[prb,aps,twocolumn,showpacs]{revtex4}
\usepackage{graphicx,amsfonts,amsmath,bm}
\newcommand{\kx}{|\mathrm{X}\rangle}
\newcommand{\ky}{|\mathrm{Y}\rangle}
\newcommand{\kup}{|\sigma_{+}\rangle}
\newcommand{\kdn}{|\sigma_{-}\rangle}
\newcommand{\kbi}{|\mathrm{XX}\rangle}
\newcommand{\kg}{|\mathrm{g}\rangle}
\newcommand{\bup}{\langle\sigma_{+}|}

\newcommand{\bbi}{\langle\mathrm{XX}|}
\newcommand{\bx}{\langle\mathrm{X}|}

\newcommand{\pbi}{|\mathrm{XX}\rangle\!\langle\mathrm{XX}|}
\newcommand{\pg}{|\mathrm{g}\rangle\!\langle \mathrm{g}|}
\newcommand{\pup}{|\sigma_{+}\rangle\!\langle \sigma_{+}|}
\newcommand{\pdn}{|\sigma_{-}\rangle\!\langle \sigma_{-}|}
\newcommand{\pp}{|+\rangle\!\langle +|}
\newcommand{\px}{|\mathrm{X}\rangle\!\langle\mathrm{X}|}
\newcommand{\py}{|\mathrm{Y}\rangle\!\langle\mathrm{Y}|}
\newcommand{\kp}{|+\rangle}
\newcommand{\km}{|-\rangle}
\newcommand{\kk}{\bm{k}}
\newcommand{\bk}{b_{\bm{k}}}
\newcommand{\bkd}{b_{\bm{k}}^{\dag}}
\newcommand{\bmkd}{b_{-\bm{k}}^{\dag}}

\newcommand{\sumk}{\sum_{\bm{k}}}
\newcommand{\wk}{\omega_{\bm{k}}}

\newcommand{\gk}{g_{\bm{k}}}
\newcommand{\rl}{\rangle\!\langle}

\DeclareMathOperator{\tr}{Tr}

\begin{document}

\title{Theory of two-photon processes in quantum dots: coherent
evolution and phonon-induced dephasing}

\author{Pawe{\l} Machnikowski}
 \email{Pawel.Machnikowski@pwr.wroc.pl} 
\affiliation{Institute of Physics, Wroc{\l}aw University of
Technology, 50-370 Wroc{\l}aw, Poland}

\begin{abstract}

The paper discusses coherent control of charge and spin
states of a biexciton system in a quantum dot via coherent two-photon
transitions. 
Rabi oscillations between the ground
state of a quantum dot and the biexciton state, as well as 
oscillations between the two single exciton states 
induced by laser pulses with
different circular or linear polarizations are studied. The effect of phonon-induced
decoherence on these processes is described. System
properties and driving conditions that lead to optimal coherent control are
identified. It is shown that proper optimization allows one to
control the two-qubit biexciton system via two-photon transitions with
a high fidelity. 
\end{abstract}

\pacs{78.67.Hc,42.50.Hz,71.38.-k,03.65.Yz}

\maketitle

\section{Introduction} 

Developing efficient methods for optical control of the quantum states
of carriers confined in semiconductor nanostructures is not only
important from a purely scientific point of view but also vital for
emerging technologies. Quantum dots (QDs) have been proposed as 
an efficient source of on-demand entangled photons \cite{benson00}.
This application was experimentally realized using optically driven QD systems
\cite{stevenson06}. Demonstration of Rabi oscillations of the exciton
occupation \cite{zrenner02} and conditional 
control of the biexciton system (two confined excitons with opposite
polarizations) \cite{li03} demonstrate the feasibility of coherent
optical control of charge (orbital) degrees of freedom. Although this
suggests that implementing quantum bits on such charge states
might be possible the relatively short lifetime of charge excitations
considerably restricts the feasibility of this solution. Instead, 
spin degrees of freedom, which are stable over much longer time
scales, 
are believed to be much more promising \cite{loss98}. Here, again,
optical methods allow one to control the system on picosecond time
scales, that is, much faster than by electron spin resonance or
by electric gating using exchange interaction. A whole range of
theoretical proposals for such optical spin control schemes 
\cite{pazy03a,troiani03,economou07,gauger08} was recently followed by
an experimental demonstration \cite{berezovsky08}. It has also been
shown that entanglement generating two-qubit gates can be performed
optically on
confined spin qubits via various all-optical schemes
\cite{kolli06,gauger08b}. 

While the atomic-like discrete properties of optical transitions in
QDs allow one, in principle, to implement a wide variety of quantum
optical schemes, the solid state nature of the system introduces
various decoherence channels that cannot be ignored when designing
optical control schemes. One of the most important sources of
dephasing in coherently driven systems is the coupling to lattice
degrees of freedom (phonons) 
\cite{forstner03,machnikowski04b,krugel05}
which imposes additional constraints on the implementation of optical
control protocols \cite{calarco03}. Typically, 
due to the dynamical character of the lattice response and to the
highly structured nature of the lattice reservoir,
the requirement of
avoiding strong lattice response restricts the parameters of the control
fields to a narrow range which can only be determined by careful
modelling of the control procedure in the presence of decoherence
\cite{roszak05b,grodecka07}. 

In order to achieve fast control with high fidelity and to provide enough
flexibility for optimizing against decoherence processes, new control
schemes are sought for. This paper is devoted to a specific class
of such schemes: quantum control of a four-level biexciton
system using coherent two-photon transitions. The direct motivation
for the present study are recent experiments in which coherent
two-photon transitions between the ground state and the biexciton
state have been demonstrated \cite{flissikowski04} and full
pulse-area-dependent two-photon Rabi oscillations 
between these two states were induced \cite{stufler06}. 
Here, a complete theory of coherent two-photon processes in a
biexciton system in a QD will be presented. 
Such processes provide an additional degree of control of the
biexciton, which can be viewed as the simplest semiconductor two-qubit system. 
They can be used to create entanglement with a single laser pulse,
which is a valuable alternative to the schemes based on sequential
transitions or two-color control.
From
the spin-oriented point of view, some of the transitions to be
described here consist in a simultaneous flip of the electron and hole
spin.  
Both the coherent evolution in an optically
driven four-level system and the effect of carrier-phonon coupling
will be discussed. It will be shown that QDs with various spectral
properties (positive vs. negative biexciton shifts) and under various
driving conditions (sign and value of the frequency detuning of the
laser field) are optimal for different control schemes.

The paper is organized as follows. Section~\ref{sec:system} introduces
the model of the system under study. Next, in Sec.~\ref{sec:unpert},
the perfect system evolution (without dephasing) is discussed, first
for linearly polarized laser pulses (Sec.~\ref{sec:unpert-linear}),
then for circularly polarized ones (Sec.~\ref{sec:unpert-circ}).
Section~\ref{sec:method} outlines the numerical and analytical (perturbative)
methods used for the study of the phonon impact on the system
evolution. In Sec.~\ref{sec:results}, the system evolution including
phonon-induced dephasing is discussed for various system
configurations. Sec.~\ref{sec:concl} concludes the paper with a
summary and final remarks. In addition, the influence of the exchange
splitting of the single exciton states is discussed in the Appendix.

\section{The system}
\label{sec:system}

\begin{figure}[tb]
\begin{center}
\unitlength 1mm
\begin{picture}(80,40)(0,5)
\put(0,0){\resizebox{80mm}{!}{\includegraphics{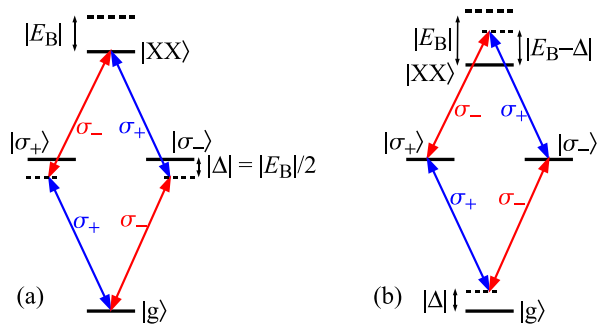}}}
\end{picture}
\end{center}
\caption{\label{fig:diags-down}The energy level structure of a confined
biexciton and the schematic presentation of the optical couplings
between the states for the biexciton Rabi oscillations (a) and for the
exciton spin flip (b).} 
\end{figure}

The system under consideration is composed of a single QD coupled to a laser beam
which may have either linear or circular polarization. 
According to the selection rules, a circularly right-polarized beam
can only induce transitions from the ground state $\kg$ to the right circularly
polarized exciton state $\kup$ and from the left circularly
polarized exciton state $\kdn$ to the biexciton state $\kbi$ (blue
arrows labeled '$\sigma_{+}$' in Fig.~\ref{fig:diags-down}). The other
two transitions (red arrows labeled '$\sigma_{-}$' in
Fig.~\ref{fig:diags-down}) are allowed for a left-polarized beam. A
linearly polarized beam is a superposition of both circularly
polarized components and 
couples both exciton states with fixed circular polarizations ($\kup,\kdn$)
to the ground and biexciton states ($\kg,\kbi$). In this way, the
system is modeled by a four-level `diamond' structure, which is a
generalization of `$\Lambda$ systems' and `V systems' studied in
quantum optics.

It is assumed that the frequency of the beam is detuned from all
the single-photon transitions, as shown in
Fig.~\ref{fig:diags-down}. For pulses of 
picosecond durations, the excited states of confined carriers are
irrelevant and may be disregarded. In most of the discussion, 
we will also neglect the exchange
interaction that couples the two circularly polarized exciton states
and turns them into a weakly split linearly polarized doublet
\cite{ivchenko97,bayer02b,hudson07}. The energy of this coupling is usually below
100~$\mu$eV and can be reduced or even cancelled by applying external fields
\cite{stevenson06b,gerardot07} or by special treatment (annealing) of
the samples \cite{tartakovskii04,langbein04c}.
In most cases, the small fine structure splitting does not
affect the system dynamics on 
the picosecond time scales relevant for the present discussion. 
However, it becomes important if the single exciton states are coupled
by long (spectrally selective) pulses, so that the fine structure
becomes spectrally resolved. This effect is discussed in the
Appendix. 
The effects of finite exciton
lifetime (of the order of 1~ns) will also be disregarded. 

Thus, the system is described by
the Hamiltonian 
\begin{displaymath}
\tilde{H}=\tilde{H}_{\mathrm{XX}}+\tilde{H}_{\mathrm{las}}
+H_{\mathrm{ph}}+H_{\mathrm{int}}.
\end{displaymath}
Here, the first term describes the four-level biexciton system,
\begin{equation*}
\tilde{H}_{\mathrm{XX}} = E(\pup+\pdn)
+ (2E +E_{\mathrm{B}})\pbi,
\end{equation*}
where $E$ is the energy of the single exciton states and
$E_{\mathrm{B}}$ is the biexciton shift (assumed
negative in the binding case, as in Fig.~\ref{fig:diags-down}).
The second term accounts for the coupling between the carrier states
and the laser beam. For a linear polarization it has the form
\begin{equation*}
\tilde{H}_{\mathrm{las}}^{({\mathrm{lin}})}
=f(t)\cos(\Omega t)\left[ (\kg+\kbi)\bx +\mathrm{H.c.}\right],
\end{equation*}
while for a circular (say, $\sigma_{+}$) polarization it reads
\begin{equation*}
\tilde{H}_{\mathrm{las}}^{({\mathrm{circ}})}
=f(t)\cos(\Omega t)\left( \kg\bup +\kdn\bbi+\mathrm{H.c.}\right),
\end{equation*}
where $f(t)$ is the envelope of the laser pulse amplitude
(which is assumed to be real), $\Omega$ is
the laser frequency and we
define `linearly polarized' exciton states $\kx,\ky$, related to the circularly
polarized states (i.e., angular momentum eigenstates) by 
$|\sigma_{\pm}\rangle=(\kx\pm i\ky)/\sqrt{2}$.
The third term in Eq.~(\ref{hX}) is the free phonon Hamiltonian,
\begin{displaymath}
H_{\mathrm{ph}}=\sumk\hbar\wk\bkd\bk,
\end{displaymath}
where $\bkd,\bk$ are creation and annihilation operators for a phonon
with a wave vector $\kk$ and $\wk$ is the corresponding frequency.
The last term describes carrier-phonon interaction,
\begin{eqnarray*}
\lefteqn{H_{\mathrm{int}}=}\\
&&(\px+\py+2\pbi)\sumk\gk^{*}\left( \bk+\bmkd \right),
\end{eqnarray*}
where the coupling constants $\gk=g_{-\bm{k}}^{*}$ 
account for the deformation potential
coupling between the longitudinal acoustic phonons and a confined
neutral exciton. This coupling dominates over the piezoelectric
coupling to the acoustic phonon branches if the
electron and hole wave functions overlap strongly
\cite{krummheuer02}. On the other hand, coupling to optical phonons is
not important in the present study since both the typical frequencies of
the system evolution and the magnitudes of the detunings
are much lower than the frequencies of the optical phonons.

For the pulse envelope, a Gaussian shape will be assumed,
\begin{displaymath}
f(t)=\frac{\hbar\theta}{\sqrt{2\pi}\tau_{0}}
e^{-\frac{1}{2}\left(\frac{t}{\tau_{0}}\right)^{2}},
\end{displaymath}
where $\tau_{0}$ is a parameter defining the pulse duration and 
\begin{displaymath}
\theta=\int_{-\infty}^{\infty}dt \frac{f(t)}{\hbar}
\end{displaymath}
is the pulse area.

Assuming 
Gaussian wave functions and neglecting the small correction resulting
from the different localization widths of electrons and holes, the
coupling constants are
given by \cite{grodecka06}
\begin{displaymath}
\gk=(\sigma_{\mathrm{e}}-\sigma_{\mathrm{h}})
\sqrt{\frac{\hbar k}{2\rho vc}}
e^{-l^{2}(k_{x}^{2}+k_{y}^{2})/4-l_{z}^{2}k_{z}^{2}/4},
\end{displaymath}
where $\sigma_{\mathrm{e,h}}$ are the deformation potential constants
for electrons and holes, $\rho$ is the crystal density, $v$ is the
normalization volume for phonon modes, $c$ is the speed of
sound, and $l,l_{z}$ are the confinement lengths in the QD plane and
along the growth direction. The values used in the calculations are
$\sigma_{\mathrm{e}}-\sigma_{\mathrm{h}}=9$~eV,
$\rho=5350$~kg/m$^{3}$, $c=5150$~m/s, $l=4.5$~nm, and $l_{z}=2$~nm.

The Hamiltonian is transformed to the `rotating frame' by the canonical
transformation defined by the unitary operator
\begin{equation*}
U=
e^{i\Omega t(\pup+\pdn +2\pbi)-i(\Delta/\hbar)t\mathbb{I}},
\end{equation*}
where $\Delta=\hbar\Omega-E$
is the detuning between the laser frequency and the
single exciton transition energy
and $\mathbb{I}$ is the identity operator.
The transformed Hamiltonian is
\begin{equation*}
H = UHU^{\dag}+i\hbar \frac{dU}{dt}U^{\dag}
= H_{\mathrm{XX}}+H_{\mathrm{las}}+H_{\mathrm{ph}}+H_{\mathrm{int}},
\end{equation*}
where
\begin{eqnarray}
\label{hX} 
H_{\mathrm{XX}} & = & \Delta\pg+(E_{\mathrm{B}}-\Delta)\pbi,\\
H_{\mathrm{las}}^{({\mathrm{lin}})}
& = & f(t)\cos(\Omega t) \nonumber \\
&&\times\left[ \left(
e^{-i\Omega t}\kg+e^{i\Omega t}\kbi\right)\bx 
+\mathrm{H.c.}\right],
\label{h-lin-full} \\
H_{\mathrm{las}}^{({\mathrm{circ}})}
& = & f(t)\cos(\Omega t) \nonumber \\
&& \times \left[ e^{-i\Omega t}\left( 
\kg\bup +\kdn\bbi \right) +\mathrm{H.c.}\right],
\label{h-circ-full}
\end{eqnarray}
and the contributions $H_{\mathrm{ph}}$ and $H_{\mathrm{int}}$ remain
unchanged. 

Since the dynamics induced by the laser pulse is slow compared to the
frequency of the optical field it is possible to treat the system in
the rotating wave approximation \cite{scully97}, as is commonly done in the
description of picosecond dynamics induced by laser fields in the
optical or near-infrared range (that is, with femtosecond oscillation
periods). One neglects in Eqs.~(\ref{h-lin-full}) and
(\ref{h-circ-full})
the very quickly oscillating terms containing 
$\exp(\pm2i\Omega t)$ which average to null over a typical time scale
characterizing the evolution of the system state. This leads to the
final formulas for the exciton-laser coupling Hamiltonians
\begin{equation}\label{h-lin}
H_{\mathrm{las}}^{({\mathrm{lin}})}
=\frac{1}{2}f(t)\left[ (\kg+\kbi)\bx +\mathrm{H.c.}\right],
\end{equation}
and
\begin{equation}\label{h-circ}
H_{\mathrm{las}}^{({\mathrm{circ}})}
=\frac{1}{2}f(t)\left( \kg\bup +\kdn\bbi+\mathrm{H.c.}\right).
\end{equation}

\section{Unperturbed evolution}
\label{sec:unpert}

In this section, various schemes of two-photon coherent control of a
biexciton system are discussed on a purely quantum-optical level, that
is, without phonon-induced dephasing. The optically induced
transitions depend on the polarization of the laser pulse. The
following two subsections are devoted to the evolution driven by
linearly and circularly polarized laser pulses, respectively.

\subsection{Linear polarization}
\label{sec:unpert-linear}

\begin{figure}[tb]
\begin{center}
\unitlength 1mm
\begin{picture}(85,30)(0,5)
\put(0,0){\resizebox{85mm}{!}{\includegraphics{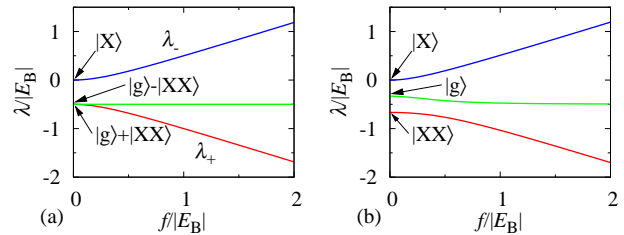}}}
\end{picture}
\end{center}
\caption[]{\label{fig:branches}The spectral branches representing the
instantaneous eigenstates of the unperturbed Hamiltonian 
$H_{\mathrm{XX}}+H_{\mathrm{las}}^{(\mathrm{(lin)})}$
as functions of the
pulse amplitude $f$ (the adiabatic parameter) for 
$\Delta=E_{\mathrm{B}}/2$ (a) and for
$\Delta=E_{\mathrm{B}}/3$ (b). In both cases, $E_{\mathrm{B}}<0$.} 
\end{figure}

Let us start with the unperturbed evolution of the system, generated by the
Hamiltonian $H_{\mathrm{XX}}+H_{\mathrm{las}}^{(\mathrm{lin})}$ 
[Eqs.~(\ref{hX}) and (\ref{h-lin})]. 
Consider the diagram of instantaneous
eigenstates of the relevant three-level Hamiltonian (excluding the
decoupled state $\ky$) as a function of the pulse amplitude $f$.
First, let us focus on the special case
of $\Delta=E_{\mathrm{B}}/2$ [Fig.~\ref{fig:branches}(a)]. 
Then
\begin{equation*}
H_{\mathrm{XX}}=\frac{\Delta}{2}(\pp+|-\rl -|),
\end{equation*}
where 
$|\pm\rangle
=(|\mathrm{g}\rangle\pm|\mathrm{XX}\rangle)/\sqrt{2}$. The driving
field [Eq.~(\ref{h-lin})] does not couple the state $\km$
to the other states, hence this state is
invariant under $H_{\mathrm{XX}}+H_{\mathrm{las}}^{(\mathrm{lin})}$. 
Another, nontrivial invariant subspace is
spanned by the states $\kx$ and $\kp$. The instantaneous eigenvalues
along the two spectral branches belonging to this subspace are 
\begin{equation}\label{lambda}
\lambda_{\pm}(t)
=\frac{\Delta}{2}
\left\{1\pm\sqrt{1+2[f(t)/\Delta]^{2}}\right\},
\end{equation}
where $\lambda_{-}$ corresponds to the branch
originating from the state $\kx$ [upper branch, blue line in
Fig.~\ref{fig:branches}(a)].  
It is clear that the two branches
are separated from each other by at least $|\Delta|$.

If the amplitude $f$ of the laser pulse
changes slowly enough in time the evolution generated by $H_{\mathrm{XX}}$
may be found using the adiabatic theorem \cite{messiah66,stufler06}.  
In such case, the states $\kx$ and $\kp$ undergo an adiabatic
evolution, with $f(t)$ playing the role of a slowly varying
parameter. At the time $t_{1}$, 
after the pulse has been switched off, the initial state is
restored with the additional dynamical phase 
\begin{equation}
\label{alfa}
\alpha_{\pm}
=-\frac{1}{\hbar}\int_{t_{0}}^{t_{1}}dt \lambda_{\pm}(t),
\end{equation}
where $t_{0}$ is the initial time (before the pulse was switched on).

\begin{figure}[tb]
\unitlength 1mm
\includegraphics[width=80mm]{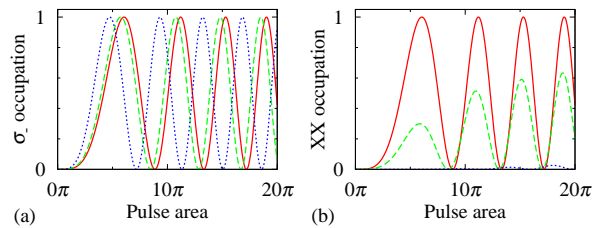}
\caption{\label{fig:2}(a) Two-photon polarization flip: 
the occupation of the state $\kdn$ as a function of the pulse area for
$\Delta=E_{\mathrm{B}}/2=-2$~meV
(red solid), $\Delta=E_{\mathrm{B}}/3$ (green dashed), and
$\Delta=E_{\mathrm{B}}/6$ 
(blue dotted). (b) Two-photon Rabi oscillations \cite{stufler06}: the
occupation of the 
biexciton state as a function of the pulse area for
$\Delta=E_{\mathrm{B}}/2=-2$~meV
(red solid), $\Delta=-1.9$~meV (green dashed), and $\Delta=-1.7$~meV
(blue dotted). In both figures, $E_{\mathrm{B}}=-4$~meV and $\tau_{0}=5$~ps.}
\end{figure}

Assume now that the system is initially prepared in the
state
\begin{displaymath}
\kup=\frac{\kx+i\ky}{\sqrt{2}}.
\end{displaymath}
The state $\ky$ is decoupled from the laser
beam and does not evolve. 
As a result, the state $\kup$ undergoes the transformation
\begin{eqnarray*}
\kup& \longrightarrow  & 
\frac{e^{i\alpha_{-}}\kx+i\ky}{\sqrt{2}}\\
& = & e^{i\alpha_{-}/2}
\left( \cos\frac{\alpha_{-}}{2}\kup
+i\sin\frac{\alpha_{-}}{2}\kdn \right).
\end{eqnarray*}
The phase $\alpha_{-}$ may be arbitrarily large. Moreover, for a fixed
pulse shape, it is a monotonous function of the pulse intensity. Thus,
by varying the pulse amplitude, 
the exciton can be coherently rotated between the two polarization
states $\kup$ and $\kdn$. The resulting occupation of the state
$\kdn$, obtained from numerical solution of the quantum evolution
equation, is shown in Fig.~\ref{fig:2}(a) (red solid line)
as a function of the pulse area $\theta$.

In the same way, since the state $\km$ evolves only trivially,
if the system is initially in the state 
\begin{displaymath}
\kg=\frac{\kp+\km}{\sqrt{2}},
\end{displaymath}
it will
undergo the transformation
\begin{displaymath}
\kg \longrightarrow 
e^{i[\alpha_{+}-\Delta(t_{1}-t_{0})/\hbar]/2} 
\left( \cos\frac{\beta}{2}\kg+i\sin\frac{\beta}{2}\kbi \right),
\end{displaymath}
where $\beta=\alpha_{+}+\Delta(t_{1}-t_{0})/(2\hbar)$.
Note that $\lambda_{+}\to\Delta$ when the laser pulse is switched off,
so that $\beta$ is in fact independent of the choice of the initial
and final time. 
Thus, when the pulse amplitude is increased the system oscillates
between the ground and biexciton states,
much like in the usual pulse-area dependent Rabi
oscillations between the ground and single exciton states, induced by a
resonant circularly polarized beam \cite{zrenner02}. The biexciton
oscillations are plotted in Fig.~\ref{fig:2}(b) (red solid line). Such
oscillations were indeed observed in an experiment \cite{stufler06}.

In spite of some qualitative similarity to the Rabi
oscillations in a two-level system,
described by the universal function $\sin^{2}(\theta/2)$,
one can clearly see essential differences. 
The two-photon oscillations are not strictly periodic,
especially for weak pulses, when the transition probability develops
very slowly. In fact, for weak pulses the occupation of the other
state grows as $\theta^{4}\sim I^{2}$, where $I$ is the pulse
intensity, as expected for a two-photon process (see
Ref.~\onlinecite{stufler06}). 
Moreover, 
it is clear from Eq.~(\ref{lambda}) that the rotation angle $\beta$
is a nonlinear
functional of the pulse envelope. Hence, contrary to the usual Rabi
oscillations, no universal area theorem exists for the final occupations.

In the general case, $\Delta\neq E_{\mathrm{B}}/2$, there are no
invariant states (apart from $\ky$) and all three states $\kg,\kx,\kbi$
give rise to three non-degenerate spectral 
branches {}[Fig.~\ref{fig:branches}(b)]. 
Now, in the adiabatic limit, the state $\kx$
evolves as previously, except for a different value of the
corresponding adiabatic
eigenvalue, which now cannot be given in a simple analytical
form. Thus, the 
exciton spin flip effect will take place also in this case, as shown
in Fig.~\ref{fig:2}(a) (green dashed and blue dotted lines). 

On the other hand,
since the states $\kg,\kbi$ now belong to non-degenerate branches, an
adiabatic evolution
starting from any of these states will end up in the
same state, up to an irrelevant global phase. Therefore, the
two-photon Rabi oscillations are suppressed when the driving field is
detuned from $\Delta=E_{\mathrm{B}}/2$, as shown in
Fig.~\ref{fig:2}(b) (green dashed and blue dotted lines). In fact, 
the symmetric detuning condition is equivalent to
the two-photon resonance between the ground and biexciton
states. If the two-photon
resonance condition $\Delta=E/2$ is not satisfied, the transition from
the ground to biexciton state is forbidden and the corresponding Rabi
oscillations are precluded. 

On the contrary, the polarization flip is always resonant and the
detuning becomes a tunable parameter. 
As will be shown in the next section, 
this additional freedom of control may be useful for
optimizing experimental parameters against phonon-induced dephasing.

\subsection{Circular polarization}
\label{sec:unpert-circ}

\begin{figure}[tb]
\begin{center}
\unitlength 1mm
\begin{picture}(85,33)(0,5)
\put(0,0){\resizebox{85mm}{!}{\includegraphics{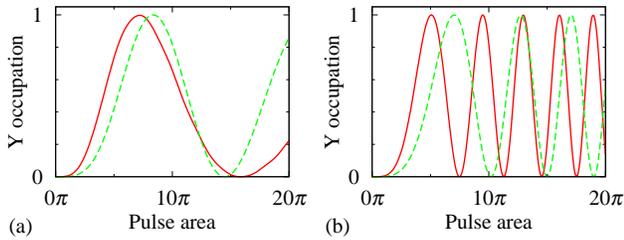}}}
\end{picture}
\end{center}
\caption{\label{fig:lin-pure}(Color online)
Two-photon rotation of the linear polarization of an exciton state
with a circularly polarized pulse: the
occupation of the $\ky$ state as a function of the pulse area for
$E_{\mathrm{B}}=-4$~meV and two values of $\Delta$: $=-3$~meV (a) and
$+2$~meV (b). Red solid lines: $\tau_{0}=5$~ps, green dashed lines:
$\tau_{0}=10$~ps. The initial state is $\kx$.} 
\end{figure}

In the case of a $\sigma_{+}$--polarized beam, the unperturbed system evolution is
generated by the Hamiltonian
$H_{\mathrm{XX}}+H_{\mathrm{las}}^{(\mathrm{circ})}$ which has two invariant
two-dimensional subspaces spanned, respectively, by the states
$\kg,\kup$ and $\kbi,\kdn$. In the adiabatic limit, each of these
four states follows its own branch of instantaneous eigenstates and
is restored after the laser is switched off, up to a phase
factor. Like in the previously discussed cases, these phase factors
can result in a nontrivial evolution if 
one starts with a system prepared in a superposition of the
states. For definiteness, let us choose the initial state
\begin{displaymath}
\kx=\frac{\kup+\kdn}{\sqrt{2}}.
\end{displaymath} 
The instantaneous eigenvalues
corresponding to the two branches departing from $\kup$ and $\kdn$ are
\begin{subequations}
\begin{eqnarray}
\lambda_{+} & = & \Delta \frac{1-\sqrt{1+[f(t)/\Delta]^{2}}}{2}, 
\label{lambda-plus} \\
\lambda_{-} & = & (E_{\mathrm{B}}-\Delta)
\frac{1-\sqrt{1+[f(t)/(E_{\mathrm{B}}-\Delta)]^{2}}}{2}.
\end{eqnarray}
\end{subequations}
The final state resulting from the adiabatic evolution is then 
\begin{displaymath}
|\psi\rangle=e^{i\frac{\alpha_{+}+\alpha_{-}}{2}}\left(
\cos\frac{\alpha_{+}-\alpha_{-}}{2}\kx
-\sin\frac{\alpha_{+}-\alpha_{-}}{2}\ky \right),
\end{displaymath}
where $\alpha_{\pm}$ is defined by Eq.~(\ref{alfa}) but with
$\lambda_{\pm}$ defined in Eqs.~(\ref{lambda-plus},b).
Thus, the polarization of the exciton state can be rotated as long as
$\alpha_{+}-\alpha_{-}\neq 0$. This condition is clearly not satisfied
at the biexciton resonance, $\Delta=E_{\mathrm{B}}/2$, but,
as long as the laser beam is detuned from this resonance,
one expects
polarization rotation to appear. This is confirmed by the numerical integration of
the evolution equation, as shown in Fig.~\ref{fig:lin-pure}. Note
that the final state is always linearly polarized, at an angle of 
$(\alpha_{+}-\alpha_{-})/2$ to the initial polarization.

\section{Phonon effects: the method}
\label{sec:method}

This section presents the general theoretical framework for
calculating the effect of the phonon-induced
decoherence on  the two-photon transitions described above.
This will be followed by the results of numerical simulations
discussed in the next section.

It is convenient to perform a unitary transformation to the phonon-dressed
exciton states \cite{machnikowski07a},
\begin{displaymath}
\mathbb{W}=\pg+(\px+\py)W+\pbi W^{2},
\end{displaymath}
where 
\begin{displaymath}
W=\exp\left[ -\sumk \frac{\gk^{*}}{\hbar\wk} ( \bk-\bmkd ) \right].
\end{displaymath}
Note that this transformation conserves the occupations of the basis
states. Working in the dressed basis physically corresponds to the
assumption that the initial state was prepared some time (a few
ps) before the operation and it is therefore surrounded by the
polaron-like coherent phonon field. Upon this transformation and
expansion to the leading order in the coupling constants, the
Hamiltonian may be written as 
$H'=\mathbb{W}H\mathbb{W}^{\dag}=H_{0}+V$, where the the first
component in the two cases of linear and circular beam polarization
is, respectively,
\begin{eqnarray*}
H_{0}^{\mathrm{(lin)}} & = & 
\Delta\pg+H_{\mathrm{ph}}+
(E_{\mathrm{B}}-\Delta)\pbi \\
&&+\frac{w}{2}f(t) \left[ (\kg+\kbi)\bx +\mathrm{H.c.}\right]
\end{eqnarray*}
and 
\begin{eqnarray*}
H_{0}^{\mathrm{(circ)}} & = & \Delta\pg+H_{\mathrm{ph}}
+(E_{\mathrm{B}}-\Delta)\pbi\\
&&+\frac{w}{2}f(t) \left( \kg\bup +\kdn\bbi+\mathrm{H.c.}\right),
\end{eqnarray*}
while the second term, describing the interaction with the phonon
reservoir, can be written as $V=S^{\mathrm{(lin/circ)}}\otimes R$, where 
\begin{displaymath}
S^{\mathrm{(lin)}}=-\frac{i}{\sqrt{2}}f(t) \km\!\bx +\mathrm{H.c.},
\end{displaymath}
\begin{displaymath}
S^{\mathrm{(circ)}}=-\frac{i}{\sqrt{2}}f(t) 
\kg\!\bup+\kdn\!\bbi +\mathrm{H.c.},
\end{displaymath}
and
\begin{displaymath}
R=i\sumk\frac{\gk^{*}}{\hbar\wk}\left( \bk-\bmkd \right).
\end{displaymath}
Here,
\begin{equation*}
w=1-\frac{1}{2}\sumk\left|\frac{\gk}{\hbar\wk}\right|^{2}
(2n_{\bm{k}}+1),
\end{equation*}
where $n_{\bm{k}}$ is the phonon occupation number,
accounts for the pho\-non-induced renormalization of the pulse amplitude
in the slow driving limit \cite{krugel05,vagov06}.
We neglect the phonon-induced energy shifts which
are very small for acoustic phonons. 

The evolution of the reduced density matrix of the biexciton subsystem
is found by numerically solving the time-convolutionless (TCL)
evolution equation \cite{breuer02} for the density matrix in the
interaction picture with respect to $H_{0}$,
\begin{equation}\label{tcl}
\dot{\rho}(t)=
-\int_{0}^{t}d\tau\tr_{\mathrm{ph}}\left[ 
V(t),\left[ V(\tau),\rho(t)\otimes\rho_{\mathrm{ph}} 
\right]  \right],
\end{equation}
where $V(t)$ is the interaction Hamiltonian $V$
in the interaction picture with respect to $H_{0}$,
$\rho_{\mathrm{ph}}$ is the phonon density matrix at the thermal
equilibrium, and $\tr_{\mathrm{ph}}$ denotes the partial trace over the
phonon degrees of freedom.

The quantitative results presented in the following section
will be obtained from a numerical solution to Eq.~(\ref{tcl}). However, much
additional insight can be gained from a perturbative approximation to
this equation 
\cite{alicki02a,alicki04a,krugel05} and from the spectral
interpretation it provides \cite{machnikowski04b,machnikowski05d}. Let
us note that, as long as the perturbation to the system evolution
remains small, the reduced density 
matrix in the interaction picture differs little from its initial
value. Hence, $\rho(t)$ on the right-hand side of Eq.~(\ref{tcl}) may
be replaced by its initial value $\rho_{0}=|\psi_{0}\rl\psi_{0}|$,
where we assume that the initial system state is pure and represented
by the state vector $|\psi_{0}\rangle$. Upon integration,
Eq.~(\ref{tcl}) then yields
\begin{equation}\label{pert}
\rho(t)=
\rho_{0}-\int_{0}^{t}d\tau\int_{0}^{\tau}d\tau'\tr_{\mathrm{ph}}\left[ 
V(\tau),\left[ V(\tau'),\rho_{0}\otimes\rho_{\mathrm{ph}} 
\right]  \right].
\end{equation}

As a simple measure of the phonon-induced
perturbation we will use the probability that the system in its actual
state $U_{0}\rho(t)U_{0}^{\dag}$ will be found in the desired,
unperturbed state $U_{0}|\psi_{0}\rangle$ (here, $U_{0}$ is the
unperturbed evolution generated by $H_{0}$). This is equal to 
$\mathcal{F}^{2}=\langle\psi_{0}|\rho(t)|\psi_{0}\rangle$, where
$\mathcal{F}$ is the fidelity for the special case of a pure state
\cite{nielsen00}. It is convenient to write 
$\mathcal{F}^{2}=1-\delta$, where $\delta$ is referred to as the error
of a quantum control operation.
A physically transparent and meaningful formula for
$\delta$ is obtained by defining the frequency-dependent operator
\begin{displaymath}
Y(\omega)=\frac{1}{\hbar}\int_{0}^{t}d\tau S(\tau)e^{i\omega\tau}
\end{displaymath}
and the phonon spectral density
\begin{displaymath}
R(\omega)=\frac{\omega^{2}}{2\pi}
\int_{-\infty}^{\infty}dt\langle R(t)R\rangle e^{i\omega t},
\end{displaymath}
where $S(t)$ and $R(t)$ denote the operators in the interaction
picture with respect to $H_{0}$.
Then, one can write
\begin{equation}\label{error}
\delta=\int_{-\infty}^{\infty}d\omega\frac{R(\omega)}{\omega^{2}}
\sum_{l}S_{l}(\omega),
\end{equation}
where the spectral functions 
\begin{equation}
\label{S-channles}
S_{l}(\omega)=|\langle\psi_{l}|Y(\omega)|\psi_{0}\rangle|^{2}
\end{equation}
can be identified with different decoherence channels and the
summation over $l$ runs through all states $|\psi_{l}\rangle$ orthogonal to
$|\psi_{0}\rangle$ (see Refs.~\onlinecite{roszak05b,grodecka07} for
details). 

\section{Phonon effects: Results}
\label{sec:results}

This section is devoted to the system evolution for different
structures of the biexciton spectrum and under various driving
conditions. In the first two subsections we will study the phonon
impact on the biexciton Rabi oscillations and exciton spin flip
induced by a linearly polarized laser field in a system with a negative
biexciton shift (bound biexciton). The third subsection contains a
discussion of the same processes in a system with antibound biexcitons
(a positive biexciton shift) and the last one deals with phonon-induced
dephasing in the two-photon
rotation of the linear exciton polarization induced by a circularly
polarized laser light.

\subsection{Biexciton Rabi oscillations for bound biexcitons}
\label{sec:results-rabi}

\begin{figure}[tb]
\begin{center}
\unitlength 1mm
\begin{picture}(85,33)(0,5)
\put(0,0){\resizebox{85mm}{!}{\includegraphics{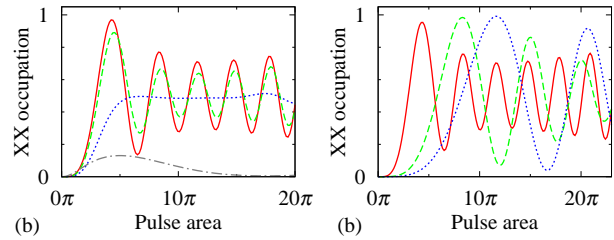}}}
\end{picture}
\end{center}
\caption{\label{fig:rabi-down}(Color online)
The two-photon Rabi oscillations in the presence
of phonon-induced perturbation. (a) The dependence of the damping on
temperature for $\tau_{0}=5$~ps: $T=0$ (red solid), $T=10$~K (green dashed), and
$T=40$~K (blue dotted). The grey dash-dotted line shows the joint occupation of 
the single exciton states due to real transitions at $T=40$~K. (b) The
dependence on the pulse duration at $T=4$~K: $\tau_{0}=5$~ps (red
solid), $\tau_{0}=20$~ps (green dashed) and $\tau_{0}=40$~ps (blue dotted).} 
\end{figure}

Let us start the discussion with the case of biexciton oscillations in
a system with a negative biexciton shift,
$E_{\mathrm{B}}<0$ (bound biexciton). In this case, the laser beam is
detuned down from the 
transition to single exciton states, as shown in
Fig.~\ref{fig:diags-down}(a).  
Therefore, one can expect that real phonon-assisted transitions to these states will
be suppressed at low enough temperatures and the decoherence will be
dominated by pure dephasing, like in the two-level 
case \cite{machnikowski04b,krugel05}. Indeed, the oscillations presented in
Fig.~\ref{fig:rabi-down}(a) show symmetric
damping up to temperatures of several Kelvins.
Only when $\Delta\sim k_{\mathrm{B}}T$,
the phonon-assisted transition up to the single exciton state $\kx$ is
possible. As a result, the damping becomes much stronger and biased
towards lower biexciton occupations, which is accompanied by a growing
single exciton component, as shown by the grey dash-dotted line
in Fig.~\ref{fig:rabi-down}(a), corresponding to $T=40$~K.

The dynamical pure dephasing effect is related to the lattice
relaxation after a non-adiabatic (with respect to the lattice response
times) change of the charge state \cite{jacak03b} which correlates the
exciton system with the phonon reservoir, generating a kind of ``which
path'' information in the macroscopic environment \cite{roszak06b}. 
Therefore, this kind of decoherence decreases when the
optically driven system evolution becomes slower
\cite{alicki04a}. This behavior is clearly visible in
Fig.~\ref{fig:rabi-down}(b). 

\begin{figure}[tb]
\begin{center}
\unitlength 1mm
\begin{picture}(85,33)(0,5)
\put(0,0){\resizebox{85mm}{!}{\includegraphics{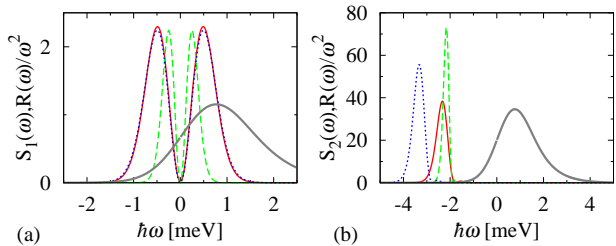}}}
\end{picture}
\end{center}
\caption{\label{fig:Sw-down}(Color online)
The spectral density of the phonon reservoir at $T=4$~K (thick grey
lines on both plots, arbitrarily scaled) and the spectral functions
$S_{1}$ (a) and $S_{2}$ (b). 
Red solid lines: $E_{\mathrm{B}}=-4$~meV, $\tau_{0}=5$~ps; 
green dashed lines: $E_{\mathrm{B}}=-4$~meV, $\tau_{0}=10$~ps;
blue dotted lines: $E_{\mathrm{B}}=-6$~meV, $\tau_{0}=5$~ps. Here
always $\Delta=E_{\mathrm{B}}/2$.}  
\end{figure}

The properties of the two contributions to dephasing (pure dephasing and real
transitions) may be conveniently studied using the approximate
perturbative formula (\ref{pert}). Fig.~\ref{fig:Sw-down} shows the
phonon spectral density $R(\omega)/\omega^{2}$ at $T=4$~K and the two spectral
functions $S_{i}(\omega)$, with $|\psi_{1}\rangle=\kbi$ and 
$|\psi_{2}\rangle=\kx$, for two different pulse durations. 
In all these plots, the nominal pulse area was tuned to achieve a complete Rabi
flop [the first maximum in Fig.~\ref{fig:2}(b)].
One can see that the first function is always positioned around the zero
frequency and its area (hence its overlap with $R(\omega)/\omega^{2}$)
decreases for slower
driving. Therefore, this spectral function may be identified with the
pure dephasing effect which decreases for slower driving.
The shape of this function does not change considerably when
the value of the biexciton shift $E_{\mathrm{B}}$ is altered.
On the other hand, the second spectral function is located in the negative
frequency area, around $\hbar\omega\approx\Delta$. This allows us to
relate this function to the real phonon-assisted transitions to the
single exciton state $\kx$. Since the phonon spectral density at large
negative frequencies is small at low temperatures this decoherence
channel is of minor importance in a system with a large enough
biexciton shift and at sufficiently low temperatures. 

\begin{figure}[tb]
\begin{center}
\unitlength 1mm
\begin{picture}(85,33)(0,5)
\put(0,0){\resizebox{85mm}{!}{\includegraphics{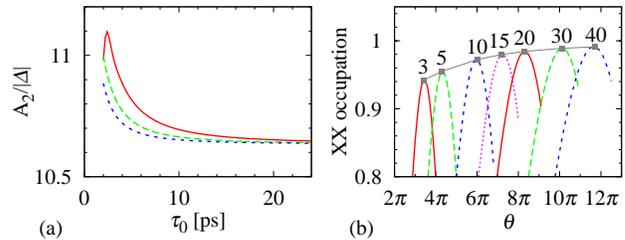}}}
\end{picture}
\end{center}
\caption{\label{fig:maxflip-down}(Color online)
(a) The value of $A_{2}/|\Delta|$ for $E_{\mathrm{B}}=-2$~meV (red
solid), $E_{\mathrm{B}}=-6$~meV (green dashed), and
$E_{\mathrm{B}}=-8$~meV (blue dotted). Here always
$\Delta=E_{\mathrm{B}}/2$. (b) 
The first maximum of the biexciton occupation for 
$\Delta=E_{\mathrm{B}}/2=-2$~meV, $T=4$~K, and pulse durations
$\tau_{0}$ as indicated. The grey line is added to guide the eye
through the maxima of the plotted curves.} 
\end{figure}

One notes in Fig.~\ref{fig:Sw-down}(b) that the shape of this spectral
feature depends both on $\Delta$ and on $\tau_{0}$. 
In order to study this real transition
process more quantitatively let us note that for a localized spectral function
$S_{2}(\omega)$, the corresponding contribution to the error may be
approximated by $\delta_{2}\approx \hbar A_{2}R(\Delta)/\Delta^{2}$, where
\begin{displaymath}
A_{2}=\hbar\int_{-\infty}^{\infty}d\omega S_{2}(\omega).
\end{displaymath}
The values of $A_{2}$ as a function of the pulse duration $\tau_{0}$ are plotted
in Fig.~\ref{fig:maxflip-down}(a). Again, for each $\tau_{0}$, the
pulse area is adjusted in order to achieve a complete Rabi flop [the
first maximum of the biexciton occupation in
Fig.~\ref{fig:2}(b)], so that all the points correspond to the same
transformation of the system state. The overall probability of a transition to
the single exciton state depends not only on the duration of the process
but also on the occupation of the single exciton state during the
evolution. The latter decreases for longer and weaker pulses. It turns
out that the two effects compensate each other almost exactly, leading
to a nearly constant value of $A_{2}$ as $\tau_{0}$ is varied. 
This means that the error
resulting from the real transition process is constant and,
consequently, the biexciton occupation should be achieved with a
constant (independent of $\tau_{0}$) accuracy as soon as the evolution
is slow enough for the pure dephasing to be negligible. This is
confirmed by the numerical solution of the TCL equation (\ref{tcl}),
presented in  
Fig.~\ref{fig:maxflip-down}(b). An additional property, clearly seen in
Fig.~\ref{fig:maxflip-down}(a), is that the
value of $A_{2}$ is nearly exactly proportional to the detuning
$\Delta$, with just a small variation (a few per cent) for very short pulses.

\subsection{Exciton spin flip}
\label{sec:spinflip}

\begin{figure}[tb]
\begin{center}
\unitlength 1mm
\begin{picture}(85,33)(0,5)
\put(0,0){\resizebox{85mm}{!}{\includegraphics{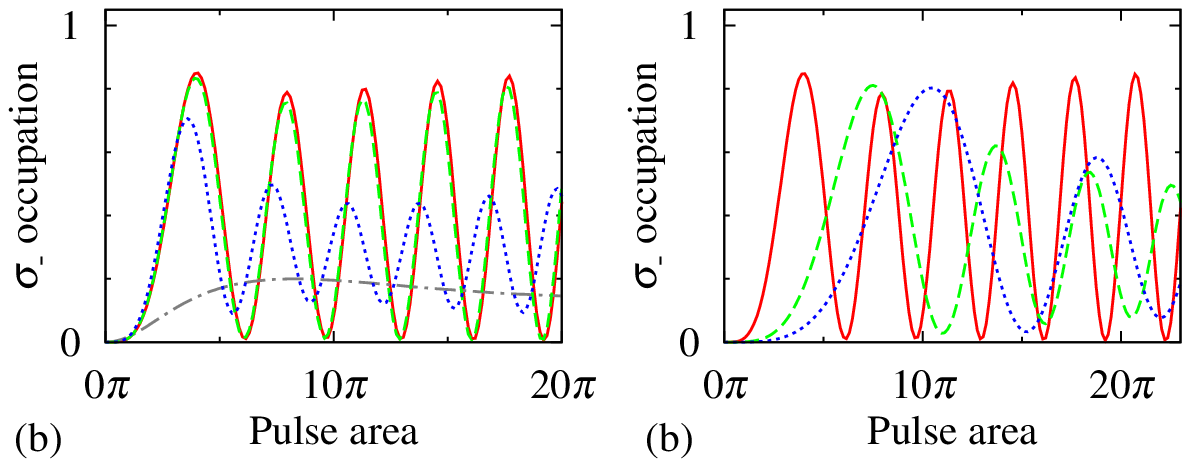}}}
\end{picture}
\end{center}
\caption{\label{fig:spinflip-down}(Color online)
The exciton spin oscillations in the presence
of phonon-induced perturbation for $E_{\mathrm{B}}=2\Delta=-4$~meV. 
(a) The dependence of the damping on
temperature for $\tau_{0}=5$~ps: $T=0$ (red solid), $T=10$~K (green dashed), and
$T=40$~K (blue dotted). The grey dash-dotted line shows the joint occupation of 
the single exciton states due to real transitions at $T=0$~K. (b) The
dependence on the pulse duration at $T=4$~K: $\tau_{0}=5$~ps (red
solid), $\tau_{0}=20$~ps (green dashed) and $\tau_{0}=40$~ps (blue dotted).} 
\end{figure}

The spectral relations are different in the case of a two-photon
spin flip. Now, for a negative biexciton energy, the laser
frequency is always detuned up from at least one of the transitions
[Fig.~\ref{fig:diags-down}(b)]. For the
detuning as in the previous case, both transitions involve
phonon emission and are therefore expected to cause much decoherence 
even at low temperatures. Indeed, the oscillations of the exciton spin
orientation shown in Fig.~\ref{fig:spinflip-down}(a) are damped in an asymmetric
way (biased towards lower occupations), which is a signature of
occupation leakage out of the two-dimensional single exciton
subspace. 
Correspondingly, single exciton occupation becomes non-zero even at
zero temperature.
Moreover, the degree of decoherence does not decrease with
a growing pulse duration [Fig.~\ref{fig:spinflip-down}(b)], again showing
that the Markovian real transition processes are a non-negligible factor. 

\begin{figure}[tb]
\begin{center}
\unitlength 1mm
\begin{picture}(85,33)(0,5)
\put(0,0){\resizebox{85mm}{!}{\includegraphics{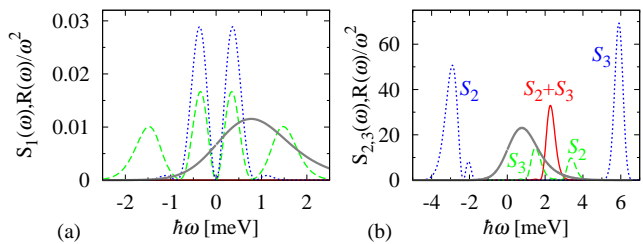}}}
\end{picture}
\end{center}
\caption{\label{fig:spin-Sw-down}(Color online)
The spectral density of the phonon reservoir at $T=4$~K (thick grey
lines on both 
plots, arbitrarily scaled) and the spectral functions $S_{1}$ (a) and
$S_{2},S_{3}$ (b), for $\Delta=-2$~meV (solid red), $\Delta=-3$~meV
(green dashed), and $\Delta=+2$~meV (blue dotted). In all the cases,
$E_{\mathrm{B}}=-4$~meV and $\tau_{0}=5$~ps.}   
\end{figure}

Interestingly, comparing Fig.~\ref{fig:spinflip-down} with 
Fig.~\ref{fig:rabi-down} one can see that
the overall dephasing of the two-photon spin flip is weaker than that
of the biexciton oscillations, except for the first maximum at low
temperatures. The reason for this is that the carrier-phonon
interaction is insensitive to the spin orientation, hence there is no
phonon-response to the transition between the two single exciton
states. There could be some perturbation resulting from the occupation
of the other two states during the evolution but in the case of
symmetric detuning, $\Delta=E_{\mathrm{B}}/2$, both these occupations
are equal and the average charge confined in the dot remains constant,
which precludes any phonon response that might lead to pure
dephasing. The spectral function related to the pure dephasing process
is again $S_{1}(\omega)$, corresponding to $|\psi_{1}\rangle=\kup$ in
Eq.~(\ref{S-channles}). In the present case, this function remains
null. This changes
slightly if the detunings from the ground and biexciton states are not
equal and, consequently, there is a charge variation during the
evolution. However, as shown in Fig.~\ref{fig:spin-Sw-down}, even in
this case the 
spectral function $S_{1}(\omega)$ is many orders of magnitude smaller
than in the case of biexciton oscillations 
[compare Fig.~\ref{fig:spin-Sw-down}(a) with Fig.~\ref{fig:Sw-down}(a)].

Phonon-assisted real transitions to the ground and biexciton states
are now described by two spectral functions $S_{2}(\omega)$ and 
$S_{3}(\omega)$,
corresponding to $|\psi_{2}\rangle=\kg$ and $|\psi_{3}\rangle=\kbi$ in
Eq.~(\ref{S-channles}), respectively. As can be seen in
Fig.~\ref{fig:spin-Sw-down}(b), these spectral features shift
according to the detuning between the laser frequency and the relevant states.

\begin{figure}[tb]
\begin{center}
\unitlength 1mm
\begin{picture}(85,33)(0,5)
\put(5,0){\resizebox{30mm}{!}{\includegraphics{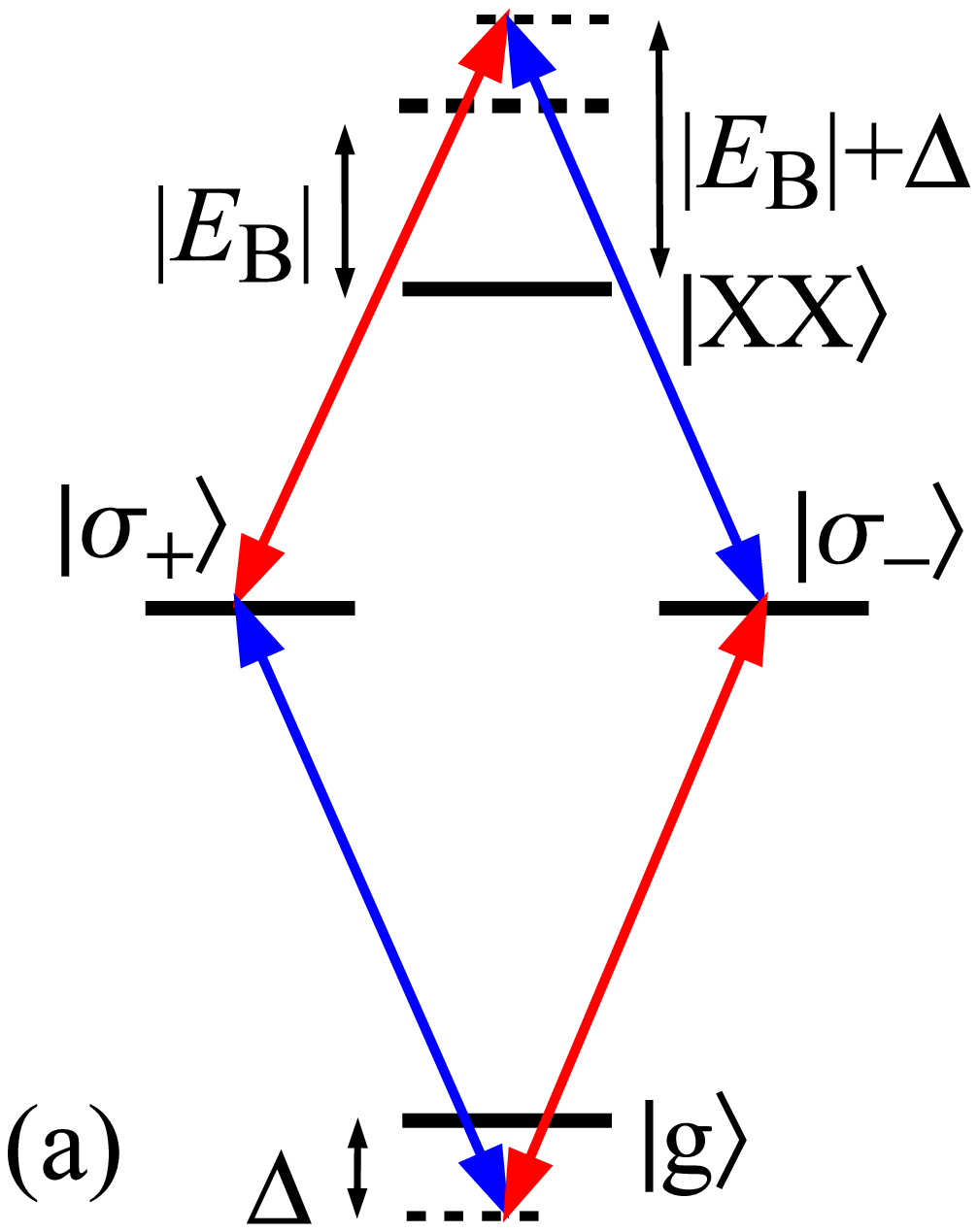}}}
\put(43,0){\resizebox{42mm}{!}{\includegraphics{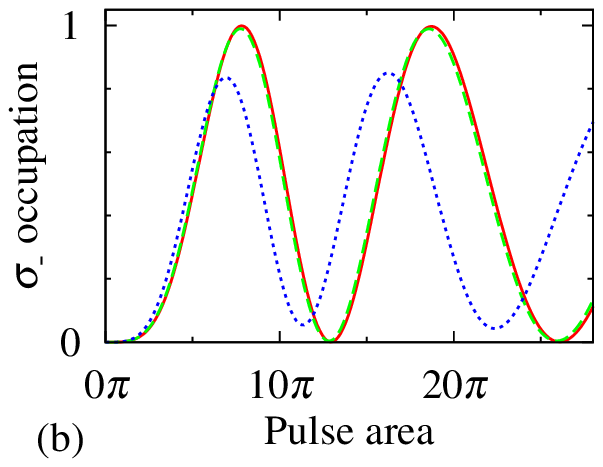}}}
\end{picture}
\end{center}
\caption{\label{fig:spinflip-up}(Color online)
(a) The diagram of energy levels and laser frequencies in the case of
$\Delta>0$, $E_{\mathrm{B}}<0$. (b) The two-photon exciton spin oscillations
in the presence of phonon-induced perturbation for $\Delta>0$,
$E_{\mathrm{B}}<0$ for $\tau_{0}=5$~ps: $T=0$ (red solid), $T=10$~K
(green dashed), and $T=40$~K (blue dotted).} 
\end{figure}

The fact that the spin flip does not require the two-photon resonance
allows one to use the detuning $\Delta$ as a free parameter to
optimize the control conditions against the phonon-induced real
transitions. The most favorable choice is to set $\Delta>0$ or
$\Delta<E_{\mathrm{B}}$ (still assuming $E_{\mathrm{B}}<0$). Then, one
of the transitions is detuned down from resonance and involves phonon
absorption which makes it ineffective at low enough temperatures [see
Fig.~\ref{fig:spinflip-up}(a)]. 
The other transition is detuned up but the detuning can be
chosen large enough to be brought beyond the cutoff of the phonon
spectral density, as shown in Fig.~\ref{fig:spin-Sw-down}(b) (dotted
blue lines). 
The simulations based on the
TCL equation~(\ref{tcl}), presented in Fig.~\ref{fig:spinflip-up}(b), show
that spin flipping 
with fidelities of the order of $10^{-3}$ is possible under such
conditions. Note that this is achieved with a relatively short pulse
so that the effect of the radiative decay during the operation is also
small (of the order of $1\%$).

\subsection{Antibound biexciton}
\label{sec:results-antibound}

\begin{figure}[tb]
\begin{center}
\unitlength 1mm
\begin{picture}(85,33)(0,5)
\put(0,0){\resizebox{85mm}{!}{\includegraphics{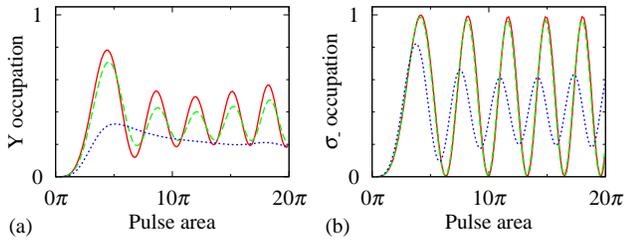}}}
\end{picture}
\end{center}
\caption{\label{fig:antibound}(Color online)
Dephasing of two-photon processes in a system with $E_{\mathrm{B}}>0$
for $\tau_{0}=5$~ps:
(a) two-photon Rabi oscillations, starting from the ground state; 
(b) two-photon spin flip, starting from the $\kup$ state. 
Red solid: $T=4$~K, green dashed: $T=10$~K, blue dotted: $T=40$~K.} 
\end{figure}

Although the unperturbed evolution for $E_{\mathrm{B}}>0$ (antibound
biexciton) does not differ from that discussed above, this is no more
true for the actual system kinetics in the presence of phonons. The
two-photon resonance condition $\Delta=E_{\mathrm{B}}/2$ that must be
satisfied for the two-photon Rabi oscillations to occur now means
that the laser beam is detuned up from the single exciton transition.
Unless this detuning is large enough (which would require large positive
biexciton shifts) phonon-assisted transitions to the single exciton
states may now take place, involving an emission of a phonon and
therefore contributing to dephasing even at low temperatures. This can
be seen in Fig.~\ref{fig:antibound}(a), 
where the results of numerical simulations are
plotted at a few values of temperature. 

On the contrary, the positive biexciton shift is favorable for the
spin flip process since now both the transitions to the ground and
biexciton state are detuned down from the resonance. Hence, 
phonon-assisted transitions may be expected to contribute considerably only
at high enough temperatures. This is indeed confirmed by the
numerical solution to the TCL equation shown in
Fig.~\ref{fig:antibound}(b), where it can be
clearly seen that considerable dephasing occurs only when the
temperature increases to a few tens of Kelvins.

One should note that the strong phonon-induced dephasing that affects
the exciton spin flip process in the $E_{\mathrm{B}}<0$ case could be
avoided by detuning the laser beam strongly above the resonance, as
discussed in Sec.~\ref{sec:spinflip}. However, strong dephasing of the
biexciton Rabi oscillations in the present case cannot be dealt with
in the same way because of the resonance condition required in this
process. 

\subsection{Exciton polarization rotation}
\label{sec:results-polarization}

\begin{figure}[tb]
\begin{center}
\unitlength 1mm
\begin{picture}(85,33)(0,5)
\put(0,0){\resizebox{85mm}{!}{\includegraphics{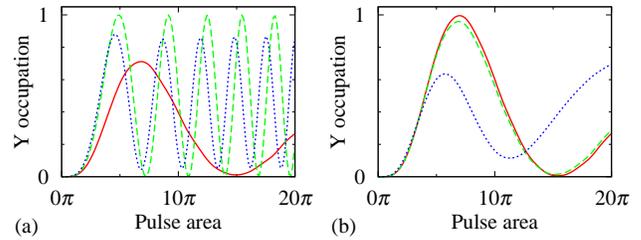}}}
\end{picture}
\end{center}
\caption{\label{fig:linear}(Color online)
Dephasing of two-photon polarization rotation for $\tau_{0}=5$~ps. 
(a) $E_{\mathrm{B}}=-4$~meV,
red solid line: $\Delta=-3$~meV, $T=4$~K; 
green dashed line: $\Delta=2$~meV, $T=4$~K; 
blue dotted line: $\Delta=2$~meV, $T=40$~K.
(b) $E_{\mathrm{B}}=4$~meV, $\Delta=3$~meV, 
red solid line: $T=4$~K; 
green dashed line: $T=10$~K; 
blue dotted line: $T=40$~K.} 
\end{figure}

Let us now discuss the possibility of rotating the linear polarization
of an exciton state via a two-photon process induced by a circularly
polarized light. The energy diagram for this process is the same as
that relevant to the spin flip and is shown in Fig.~\ref{fig:diags-down}(b).
As discussed in Sec.~\ref{sec:unpert-circ},
performing this two-photon transfer requires detuning off the
two-photon biexciton resonance, i.e., $\Delta\neq E_{\mathrm{B}}/2$.
For a system with a bound biexciton ($E_{\mathrm{B}}<0$), this means
that the detuning between the laser frequency and the transition to
either the ground state or the  biexciton state decreases and enters
the range of high 
spectral density of the phonon reservoir. This results in an increased
contribution from phonon-assisted transitions and, therefore, in 
considerable damping of the polarization oscillations even at low
temperatures, as shown in 
Fig.~\ref{fig:linear}(a). Efficient polarization rotation is only possible for large
positive detunings  [see the diagram in
Fig.~\ref{fig:spinflip-up}(a)], for the same 
reasons as discussed in Sec.~\ref{sec:spinflip}. Again, the quality of
oscillations is reduced as soon as the temperature becomes comparable
with the detuning from the ground state.

In a system with a positive biexciton shift, $E_{\mathrm{B}}>0$,
phonon-induced transitions would require a phonon absorption and are
therefore suppressed at low temperatures, as shown in
Fig.~\ref{fig:linear}(b) (red solid and green dashed lines). 
It should be noted, however, that the system evolution is relatively
slow in this regime of operations since the detunings are close to the
two-photon resonance, when the effect of the laser field completely
vanishes (see Sec.~\ref{sec:unpert-circ}). Therefore, large
pulse intensities are needed to achieve the desired transformation of
the system state.
At higher temperatures [blue dotted line in Fig.~\ref{fig:linear}(b)]
the phonon absorption 
processes contribute considerably to the dephasing of the polarization
rotation process.

\section{Conclusion}
\label{sec:concl}

Two-photon transitions driven by picosecond laser pulses detuned from
both exciton and biexciton resonances in a quantum dot open the way to
a rich variety of control schemes in a biexciton system. Using a
linearly polarized 
laser field one can induce Rabi oscillations directly between the
ground and biexciton state and coherently flip the circular
polarization of an exciton state. The former involves a two-photon
absorption or emission and requires exact two-photon
resonance. The latter can be interpreted as `photon exchange'
(absorption--emission) and can be induced in a broad range of
detunings. With a circularly polarized field, one can induce a rotation
of the linear polarization of an exciton. This process takes place
only off the two-photon resonance.
The physics beyond these two-photon processes is highlighted by
invoking the quantum-mechanical adiabatic theorem. 
If the individual excitons are treated
as separate quantum subsystems (or qubits) then these two-photon
control operations make it possible to prepare entangled states of
these subsystems with a single laser pulse. 

Already these results, restricted to the ideal evolution, confirm once
more that the biexciton system is an interesting object of investigation from
the point of view of quantum optics. They show that couplings between
all four states forming the `diamond' structure of the biexciton
levels is essential for the system dynamics even if large detuning
precludes real transitions between these levels. This suggests that
models used for three-level atomic systems (`V-systems' \cite{wang05}) may be
insufficient for a description of QD systems. 

In a real semiconductor system, the evolution is strongly affected by
phonon-induced dephasing. In addition to the pure dephasing process
appearing in an optically driven two-level exciton system, here we
deal also with real phonon-assisted transitions to the states which,
in the ideal case, should remain unoccupied. The effect of all the
dephasing channels depends on the control scheme under consideration
and on the energy level structure of the biexciton system. For
negative biexciton shifts (bound biexcitons), 
the phonon-induced decoherence is relatively weak for two-photon Rabi
oscillations and becomes stronger for the polarization flipping
process. However, the additional freedom of detuning allows one to
optimize the driving conditions against these decoherence effects also
in the latter case so that spin flip fidelities as high as $10^{-3}$
are achievable. QDs with a positive biexciton shift are less
suitable for these optical control schemes. In conrast, they are
favorable for the two-photon rotation of the linear polarization:
in this case, optimization of the control conditions against
phonon-induced dephasing can be much more efficient in a QD with
an antibound biexciton than in a QD with a bound biexciton. 

The study of phonon-induced dephasing shows, in addition, that
a two-photon transition between the two-single exciton states can be
performed via states with constant average occupation of the QD. This
eliminates the phonon response to the charge
evolution and allows one to avoid the resulting dephasing. With this respect,
the single-pulse two-photon control has a clear advantage over
sequential transitions.  

Out of the various processes discussed here, one (the biexciton
oscillations) has already been observed
experimentally \cite{stufler06}. 
The good agreement between the description based on the
adiabatic theorem and the experimental results shows
that the theory captures the essentials of the quantum evolution under actual
laboratory conditions. The other processes take place under similar
experimental conditions and therefore also seem to be experimentally
feasible. 

\begin{acknowledgments}
The author is grateful to A. Grodecka for reading the manuscript and
many helpful comments.
This work was supported by Grant No. N202-071-32-1513 of the
Polish MNiSW. 
\end{acknowledgments}

\begin{appendix}
\section{Fine structure splitting}

In almost all QDs, electron-hole exchange interaction splits the
single exciton states into a linearly polarized doublet of states
$\kx$ and $\ky$. The magnitude $\delta_{\mathrm{fs}}$ 
of this fine structure splitting can vary from several
$\mu$eV to about $100\,\mu$eV. This splitting has no essential effect
for the biexciton Rabi oscillations, while speaking of the
exciton polarization flip makes sense only when the circularly
polarized states can be considered well-defined, that is, when the
total duration of the experiment does not exceed
$\hbar/\delta_\mathrm{fs}$, which is of the order of 10 to 100 ps. 

In this Appendix the effect of the fine structure splitting on the
linear polarization rotation is discussed. The two linearly polarized
states are well defined in the presence of this splitting. Due to the
requirement for energy conservation, the
fidelity of this type of coherent control depends on the relation
between the pulse length and the magnitude of the splitting.

\begin{figure}[tb]
\begin{center}
\unitlength 1mm
\begin{picture}(85,33)(0,5)
\put(0,0){\resizebox{85mm}{!}{\includegraphics{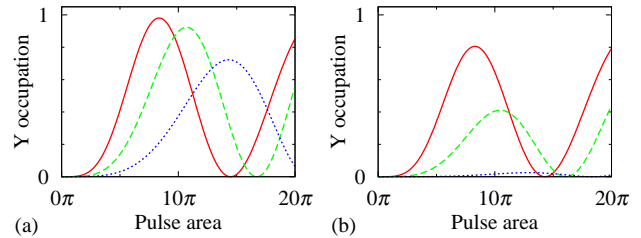}}}
\end{picture}
\end{center}
\caption{\label{fig:fine}
(Color online)
Rotation of the linear polarization of an exciton state
with a circularly polarized pulse with the fine structure splitting of
$\delta=30\,\mu$eV (a) and $\delta=100\,\mu$eV (b): the
occupation of the $\ky$ state as a function of the pulse area for
$E_{\mathrm{B}}=-4$~meV and $\Delta=-3$~meV; pulse durations
$10$~ps (red solid lines), $20$~ps (green dashed
lines) and $40$~ps (blue dotted lines).} 
\end{figure}

The additional contribution to the Hamiltonian, describing the fine
structure splitting, is 
\begin{equation*}
H_{\mathrm{fs}}=\frac{\delta_{\mathrm{fs}}}{2}(\px-\py).
\end{equation*}
The results of simulations of the system dynamics (without dephasing)
are presented in Fig.~\ref{fig:fine} for two values of the fine structure
splitting $\delta_{\mathrm{fs}}$. 
If the pulse is long and, therefore,
spectrally narrow, the transition between the two non-degenerate
states becomes forbidden by energy conservation. As can be seen, for a
typical value of $\delta_{\mathrm{fs}}=30\,\mu$eV, the effect of
lifting the degeneracy becomes important only for pulse durations of a
few tens of picoseconds.

\end{appendix}


\end{document}